\documentclass[final]{cvpr}

\usepackage{times}
\usepackage{epsfig}
\usepackage{graphicx}
\usepackage{amsmath}
\usepackage{amssymb}

\usepackage{xcolor}
\usepackage{algorithm}
\usepackage[noend]{algpseudocode}

\usepackage{subfigure}
 
\usepackage{algorithm}
\usepackage[noend]{algpseudocode}

\usepackage[pagebackref=true,breaklinks=true,colorlinks,bookmarks=false]{hyperref}

\def\cvprPaperID{****} 
\def\confYear{CVPR 2021}

\usepackage{lipsum}

\graphicspath{{./pics/}}
\def \name{\texttt{SurFree}\xspace}
\def \geoda{\texttt{GeoDA}~\cite{Rahmati:2020aa}\xspace}
\def \qeba{\texttt{QEBA}~\cite{Li:2020aa}\xspace}
\def \hsja{\texttt{HSJA}~\cite{Chen:2020aa}\xspace}

\begin{document}

\title{\name: a fast surrogate-free black-box attack}

\author{Thibault Maho\\
Univ. Rennes, Inria, CNRS, \\
IRISA, Rennes, France\\
{\tt\small thibault.maho@inria.fr}
\and
Teddy Furon\\
Univ. Rennes, Inria, CNRS, \\
IRISA, Rennes, France\\
	{\tt\small teddy.furon@inria.fr}
\and
Erwan Le Merrer\\
Univ. Rennes, Inria, CNRS, \\
IRISA, Rennes, France\\
	{\tt\small erwan.le-merrer@inria.fr}
}
\maketitle

\begin{abstract}

  Machine learning classifiers are critically prone to evasion attacks.
  Adversarial examples are slightly modified inputs that are
  then misclassified, while remaining perceptively close to their
  originals. Last couple of years have witnessed a striking
  decrease in the amount of queries a black box attack submits to the
  target classifier, in order to forge adversarials. This
  particularly concerns the black box \textit{score-based} setup,
  where the attacker has access to top predicted probabilites:
  the amount of queries went from to millions of to less than a thousand.
  
  This paper presents \name, a
  geometrical approach that achieves a similar drastic reduction in
  the amount of queries in the hardest setup: black box
  \textit{decision-based} attacks (only the top-$1$ label is
  available).  We first highlight that the most recent attacks in that
  setup, \hsja, \qeba and \geoda all perform costly gradient surrogate
  estimations. \name proposes to bypass these, by instead focusing on
  careful trials along diverse directions, guided by precise indications
  of geometrical properties of the classifier decision boundaries. We
  motivate this geometric approach before performing a
  head-to-head comparison with previous attacks with the amount of
  queries as a first class citizen. We exhibit a faster
  distortion decay under low query amounts (few hundreds to a
  thousand), while remaining competitive at higher query budgets.\footnote{The code of \name is available at \url{https://github.com/t-maho/SurFree} }
  
\end{abstract}

\section{Introduction}
\label{sec:Introduction}

The literature on adversarial examples is divided into two shares,
depending on the threat model: either the attacker has full knowledge
of the target classifier \cite{Carlini:2017aa,Rony:2019aa,Madry:2018aa} (white-box setting)
or she/he has an unrestricted query access to the unknown classifier
\cite{Narodytska:2017aa,Brendel:2018aa,Li:2020aa,Rahmati:2020aa,Chen:2020aa,Zhao:2020aa,Ilyas:2019aa,Tu:2019aa,Guo:2019aa,Cheng:2019aa,Ilyas:2018aa,Chen:2017aa}
(black-box setting).  The latter scenario is deemed as more relevant
to gauge the intrinsic robustness of classifiers in real-world
applications (typically queried through an API).

\begin{figure}[!t]
	\centering
	\resizebox{0.807\columnwidth}{!}{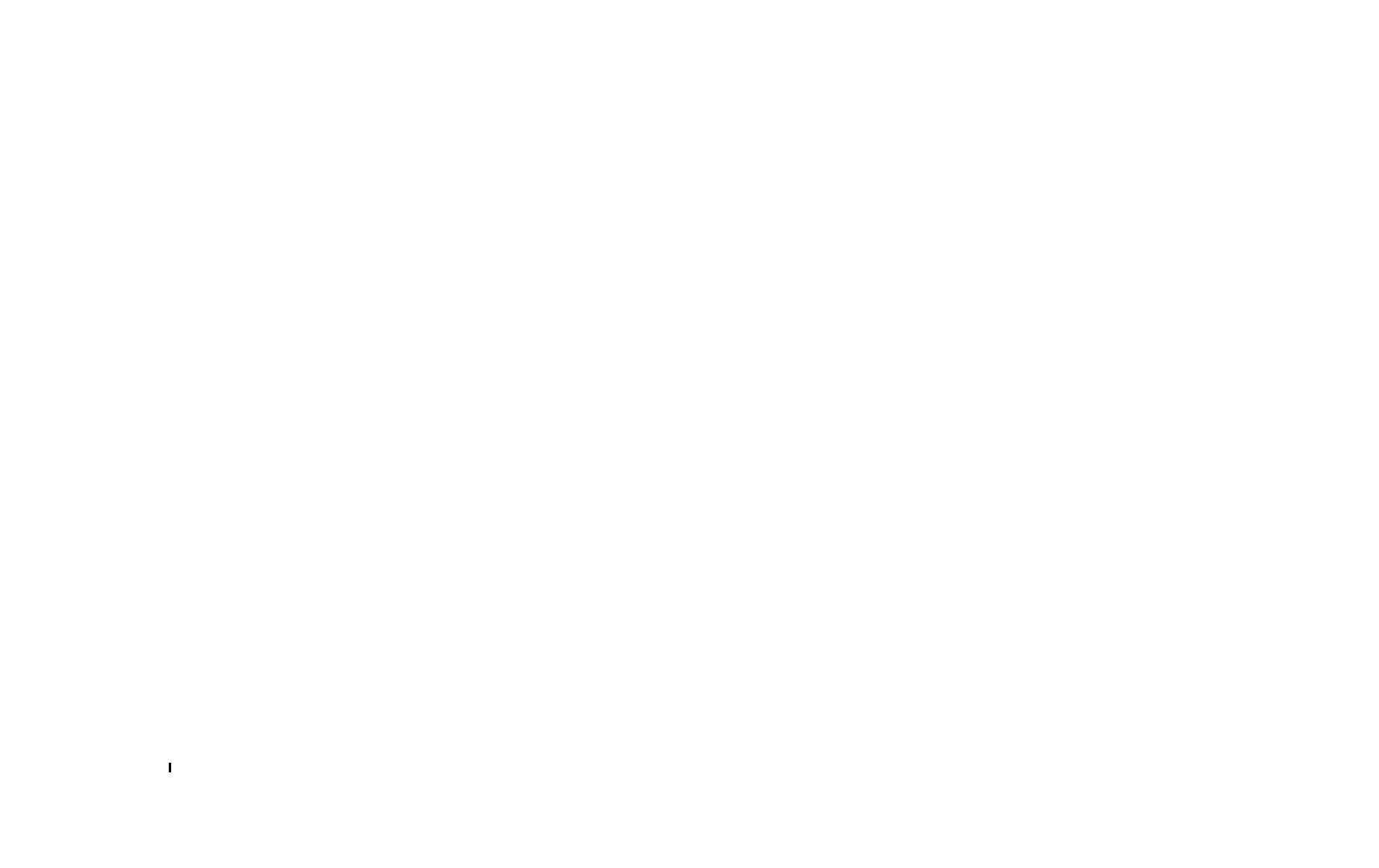}
	\includegraphics[height=4.15cm]{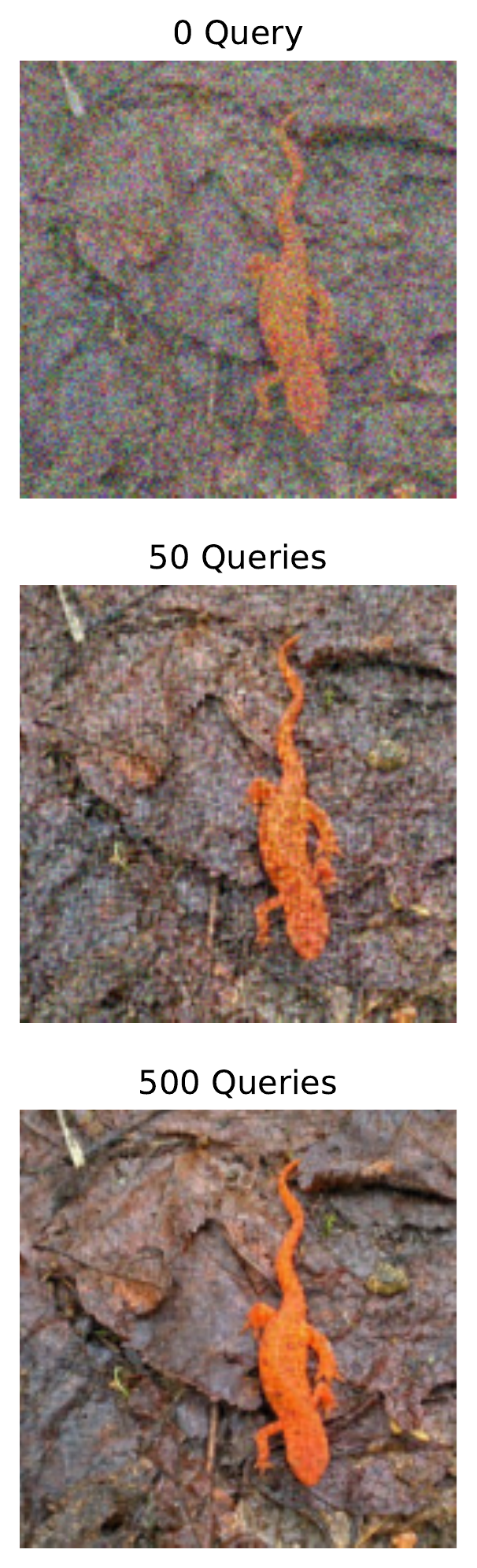}
	\caption{The perturbation distortion ($\ell_2$ norm) vs. the number of queries for image `lizard'.
		Competitor attacks waste queries to estimate a gradient surrogate resulting in plateaus of distortion.
		\label{fig:budget}}
\end{figure}

Black box attacks are iterative procedures that keep on refining the quality of an adversarial example based on the pairs of submitted input~/~observed output. 
They are coined \emph{score-based} when the attacker observes the top-$k$ predicted probabilities or \emph{decision-based} (a.k.a. hard label) when she/he only learns the top-$k$ labels ($k\geq1$).
Indeed, the latter case where the output is solely the top-1 label is the most challenging because the attacker cannot rely on
any rich information for crafting these adversarial examples.   

It is striking that black-box attacks always use substitution to replace information they are missing. 
Early black-box attacks used a surrogate model (trained from a huge number of input~/~output pairs) mimicking the targeted model~\cite{Papernot:2016aa, Papernot:2017aa}.
The attack then boils down to a white-box setting on the surrogate with the hope that the adversarial example transfers to the target classifier.   
Almost all recent score-based attacks resort to gradient estimation to compensate for the lack of back-propagation, which is the key instrument of any white-box attack~\cite{Chen:2017aa,Tu:2019aa,Ilyas:2019aa,Zhao:2020aa}.
The HopSkipJump attack (\hsja) estimates the decision boundary by an hyperplane.  
As a last example, authors in~\cite{Ilyas:2018aa} turn a decision-based setup into a score-based by probing noisy versions of an image to derive a score-like function from the top-$k$ labels. The trend is thus to substitute missing information by estimates in order to fall back to an easier setup. 

The need for faster attacks consuming fewer queries is already present in the literature.
Most notably, research works on score-based attacks managed to reduce query amount from millions of requests \cite{Ilyas:2018aa} to less than a thousand  with most recent approaches~\cite{Zhao:2020aa}.
Surprisingly, this impressive decrease has not reached comparable levels in the hard-label setup. In particular, paper~\cite{Ilyas:2018aa} questions the model surrogate approach: while a considerable amount of queries is spent for training the surrogate, not a single adversarial example is forged. Moreover, access to the target model in practice is usually not free and not unlimited\footnote{see \textit{e.g.}, {\scriptsize\url{https://azure.microsoft.com/en-us/pricing/details/cognitive-services/face-api/}} for  conditions.}.

Yet, this argument should challenge any substitution mechanisms.
They all consume a fair amount of queries and it is not clear whether they are worth the gain in term of distortion.
Especially, many techniques trade some query amount for an accurate
gradient estimate giving birth to good perturbation directions~\cite{Rahmati:2020aa,Chen:2020aa}.
While spending these queries, the adversarial example is not updated and the distortion stalls as illustrated in Fig.~\ref{fig:budget}.

This paper considers the query amount as a central criterion. It presents a \emph{fast} black-box decision-based attack, named \name, motivated by practical applications in which a low amount of queries is key.
Fast means that it outperforms the state of the art when it comes to the distortion of adversarials under a low query budget (as examplified in Fig.~\ref{fig:budget} with the purple curve).

The main contributions of this paper are:
\begin{itemize}
\item \name, a black box decision-based attack not using any substitution mechanism:
no surrogate of the target model,  no score reconstruction, no estimation of gradient.
It is inspired by the early works~\cite{Earl:2007aa,Brendel:2018aa}.
\item a geometrical mechanism to get the biggest distortion decrease for a given direction to be explored
under the assumption of a hyperplane boundary \cite{Fawzi_2018_CVPR}. 
\item a head to head comparison of the recent approaches with distortion as a function of query number.
\end{itemize}
Experimental results show that \name overcomes state of the art on the query amount factor (a thousand of
queries), while still remaining competitive with unlimited queries (normal scenario for competitors).

\def \DCT{\mathtt{DCT}}

\section{Related Works}
\label{sec:Related}

\subsection{Watermarking}
Digital watermarking embeds a secret and invisible mark into images.
A watermark detector is a two-class classifier checking for the presence or absence of the mark in a query image.
This community called oracle attack what we now call a black-box attack:
the attacker has the secret-keyed detector in hand as a black sealed box,
and calls it iteratively to either estimate the secret key or to remove the watermark from protected images.
This latter problem is equivalent to forging adversarial images in the decision based setting.

All the ingredients used nowadays were already present in this literature dating back to 1997~\cite{Cox:1997aa, Linnartz:1998aa}: surjection onto the boundary with binary search, estimation of the gradient at a `sensitive' point lying on the boundary, dimension reduction.
The new HopSkipJumpAttack~\cite{Chen:2020aa} is indeed very similar to the old Blind Newton Sensitivity Attack~\cite{Comesana:2006aa}.
The last work on this subject by this community~\cite{Earl:2007aa} surprisingly does not use any gradient estimate but random directions; this is also the case of the very first decision-based attack~\cite{Brendel:2018aa}.

\subsection{Black box adversarial examples}
\label{ssec:BB}
This paper operates in a black box setup; 
white box attacks \cite{Carlini:2017aa,Rony:2019aa,Madry:2018aa} are considered out of its scope.

There has been a huge improvement on the amount of queries of \emph{score-based} black box attacks.
On ImageNet, the order of magnitudes of the first attacks were of some hundreds of thousands queries for one image with a reported runtime of 20 minutes in~\cite{Chen:2017aa}.
Nowadays state-of-the-art attacks make less than one thousand calls to the classifier, thanks to advanced gradient estimators~\cite{Ilyas:2019aa} and Zero Order Optimization techniques~\cite{Tu:2019aa,Zhao:2020aa,Liu:2020aa}.  

The \emph{decision-based} attacks followed the same trend but with a factor of ten.
Brendel \etal report in the order of one million of queries for one image in one of the first black-box decision based paper~\cite[Fig.~6]{Brendel:2018aa}.
Then, the order of magnitude went down to tens of thousands~\cite[Fig.~4]{Chen:2020aa} \cite[Fig.~5]{Li:2020aa} and even some thousands in~\cite[Fig.~2]{Rahmati:2020aa}.  
No decision-based paper reports results with less than one thousand of calls on ImageNet.
This paper explores this range of query budget.

The main engine of the \emph{decision-based} attacks iterates the three following steps:
i) the surjection (find a point on the boundary),
ii) the estimation of the gradient (\ie the normal vector of the tangent hyperplane),
iii) the update of the adversarial example.
Step ii) proceeds by bombarding the model with small perturbations around the boundary point.
The main problems are the trade-off between the number of queries devoted to this task and the accuracy of the gradient estimate (see~\cite[Th.~2]{Chen:2020aa}, \cite[Lemma~2]{Rahmati:2020aa}, \cite[Th.~1]{Li:2020aa})
and the impact of this accuracy on the convergence of the attack (see~\cite[Th.~2]{Rahmati:2020aa}). In the end, \cite{Rahmati:2020aa} recommends that step ii) consumes a number of queries following a geometric sequence w.r.t. the iteration number, whereas \cite{Chen:2020aa} makes it proportional to the square root of the iteration number.
This paper follows the opposite strategy: no query is spent for a gradient estimate. 

A second track of improvement is dimension reduction restricting the perturbation to a low dimension subspace.
This a priori increases of distortion at convergence since the attacker has fewer degrees of freedom, but it indeed facilitates the estimation of the projected gradient.
The latter is more important for low query budget.
The choice of the subspace incorporates prior information: it usually corresponds to a low-frequency band (of the full $\DCT$ transform \cite{Rahmati:2020aa,Li:2020aa}) containing most critical information about the visual content of the image.
This paper shows that the block $\DCT$ yields better results.

\def \real{\mathbb{R}}
\def \x{\mathbf{x}}
\def \y{\mathbf{y}}
\def \n{\mathbf{n}}
\def \N{\mathbf{N}}
\def \t{\mathbf{v}}
\def \tt{\mathbf{t}}
\def \u{\mathbf{u}}
\def \z{\mathbf{z}}
\def \c{\mathbf{c}}
\def \dec{\mathsf{cl}}
\def \out{\mathcal{O}}
\def \V{\mathcal{V}}
\def \dout{\partial \out}
\def \Plan{\mathcal{P}}
\def \taub{\boldsymbol{\tau}}
\def \dim{D}
\newtheorem{prop}{Property}

\section{Problem statement}
\label{sec:Problem}
We introduce the following notations.
The pre-trained classifier is represented as the function $f:[0,1]^\dim\to\real^C$.
For a given input image $\x$, the final decision is the top-1 label $\dec(\x):=\arg\max_k f_k(\x)$, $f_k(\x)$ being the predicted probability of class $k$, $1\leq k\leq C$.

The attacker does not know the function $f$ and can only observe the decision $\dec(\x)$ for any image $\x$.
From an original well classified image $\x_o$, the attack is untargeted as it looks for an image $\x_a$ close to $\x_o$ and s.t. $\dec(\x_a)\neq\dec(\x_o)$. This defines the outside region $\out := \{\x\in\real^\dim: \dec(\x)\neq\dec(\x_o)\}$ and the optimal adversarial image:
\begin{equation}
\label{eq:OptDef}
\x_a^\star = \arg \min_{\x\in\out} \|\x-\x_o\|.
\end{equation}
This is a hard problem and the attack is indeed an efficient algorithm finding an approximate solution.

We assume that when knowing a point $\y \in \out$, it is possible to find a point $\x_b\in[\x_o, \y]$ that lies on the boundary denoted by $\dout$.
This is usually done by a line search in the literature \cite{Chen:2020aa,Li:2020aa,Rahmati:2020aa}.
There has been experimental evidence that the boundary is a rather smooth low curvature surface for deep neural networks \cite{NIPS2016_7ce3284b}.
This justifies that the boundary is often approximated by an hyperplane locally around a boundary point: 
in other words, locally around $\x_b\in\dout$,
there exists $\n\in\real^\dim$, $\|\n\|=1$ s.t. $\y\in\out$ if $\y^\top \n \geq \x_b^\top \n$.

\section{Our Approach}
\label{sec:Approach}
%


The study of the recent attacks \cite{Chen:2020aa, Rahmati:2020aa, Li:2020aa} under the query budget viewpoint,
reveals the presence of plateaus (see Fig.~\ref{fig:budget}).
These are due to the construction of a surrogate for gradients, and appear to be particularly costly.
Moreover, the budget allocated to gradient estimate in~\cite{Chen:2020aa} does not impact the speed of convergence:
fewer queries give less accurate gradient estimates yielding a smaller distortion decrease but at a higher rate.   
Our rationale is to set this query budget to its extreme value, \ie zero. 
We thus trade this budget for more directions investigated with the hope that their multiplication allows for a faster distorsion decrease.
We now develop this idea.

\subsection{Basic idea} 
\label{ssec:BasicIdea}
Let us assume that we know a point on the boundary: $\x_b\in\dout$.
We define $d:=\|\x_b-\x_o\|$ and $\u:=(\x_b-\x_o)/d$ so that $\|\u\| = 1$.
We restrict the search for a closer adversarial point in a random affine plane $\Plan$ of dimension 2.
This plane $\Plan$ contains the point $\x_o$ and is spanned by  vector $\u$ and a random orthogonal direction $\t\in\real^\dim$, $\|\t\|=1$, $\t^\top\u = 0$.
Note that $\x_b\in\Plan$.

In polar coordinates, we consider a point in $\Plan$ that is at a distance $d(1-\alpha)$ from $\x_o$ and makes an angle $\theta$ with $\u$:
\begin{equation}
\label{eq:Z}
\z(\alpha,\theta) = d(1-\alpha)\left(\cos(\theta)\u + \sin(\theta)\t\right) + \x_o,
\end{equation}
with $\alpha\in[0,1]$ and $\theta\in[-\pi,\pi]$. Note that $\z(0,0) = \x_b$ and $\z(1,\theta) = \x_o, \forall \theta$.
If $\z(\alpha,\theta)$ is adversarial, then the distortion decreases by $100\times\alpha\%$.

This section shows how to choose $(\alpha,\theta)$ to raise the probability of $\z(\alpha,\theta)$ being adversarial.
This study makes a clear cut with~\cite{Earl:2007aa,Brendel:2018aa}  which also consider random directions.

This section assumes that the intersection $\dout\cap\Plan$ is a line passing by $\x_b$ and with normal vector $\n\in\Plan$, $\|\n\|=1$.
Without loss of generality, $\n$ is pointing outside s.t. a point $\z\in\Plan$ is adversarial if $(\z-\x_b)^\top\n \geq 0$.
In polar coordinates, $\n := \cos(\psi)\u + \sin(\psi)\t$ with $\psi\in(-\pi/2,\pi/2)$.

The point $\z(\alpha,\theta)\in\dout\cap\Plan$ minimizing the distance from $\x_o$ is the projection of $\x_o$ onto this line, obtained for $\theta = \psi$ and $\alpha = 1-\cos(\psi)$.
The attacker can not create this optimal point because angle $\psi$ is unknown.
Note that
\begin{itemize}
\item If $\psi = 0$, then $\n = \u$, $\t^\top\n = 0$,  $(\z(\alpha,\theta) - \x_b)^\top\n = d((1-\alpha)\cos(\theta) - 1)<0$, and $\z(\alpha,\theta)$ is not adversarial.
This corresponds to the case where $\dout\cap\Plan$ is a tangent line of the circle of center $\x_o$ and radius $d$.
This implies that $\x_b$ is already optimum because it is the projection of $\x_o$ onto $\dout\cap\Plan$.
\item If $\theta=0$ and $\alpha>0$, then $(\z(\alpha,0) - \x_b)^\top\n = \alpha(\x_o - \x_b)^\top \n < 0$ because $\x_o$ is not adversarial.
Therefore, $\z(\alpha,0)$ is not adversarial.
\end{itemize}
For $\theta\neq 0$, calculation shows that $\z(\alpha,\theta)$ is adversarial if
\begin{equation}
\label{eq:CondAdv}
g_\alpha(\theta) := \left|\frac{1-(1-\alpha)\cos(\theta)}{(1-\alpha)\sin(\theta)}\right|\leq \tan(\psi)\mbox{sign}(\theta).
\end{equation}
Point $\z(\alpha,\theta)$ might be adversarial only if  $\psi$ and $\theta$ share the same sign s.t. the rhs~\eqref{eq:CondAdv} is positive.
In this case, the surprise is that~\eqref{eq:CondAdv} separates parameters $(\alpha,\theta)$ that the attacker controls from the unknown angle $\psi$. 

Minimizing $g_\alpha(\theta)$ raises the chances that~\eqref{eq:CondAdv} holds.
Its derivative cancels for $\theta = \theta^\star(\alpha) := \pm \arccos(1-\alpha)$ (according to the sign of $\psi$) so that
\begin{equation}
\label{eq:gAlpha}
g_\alpha(\theta^\star(\alpha)) = \frac{\sqrt{1-(1-\alpha)^2}}{1-\alpha} = |\tan(\theta^\star(\alpha))|.
\end{equation}
This quantity is an increasing function of $\alpha$ ranging from $0$ ($\alpha=0$) to $+\infty$ ($\alpha\to1$).
From now on, we denote by $\z^\star(\theta)$ a point created with this coupling between $\theta$ and $\alpha$, $\z^\star(\theta):=\z(1-\cos(\theta),\theta)$. 

\begin{figure}[tb]
\centering
\resizebox{\columnwidth}{!}{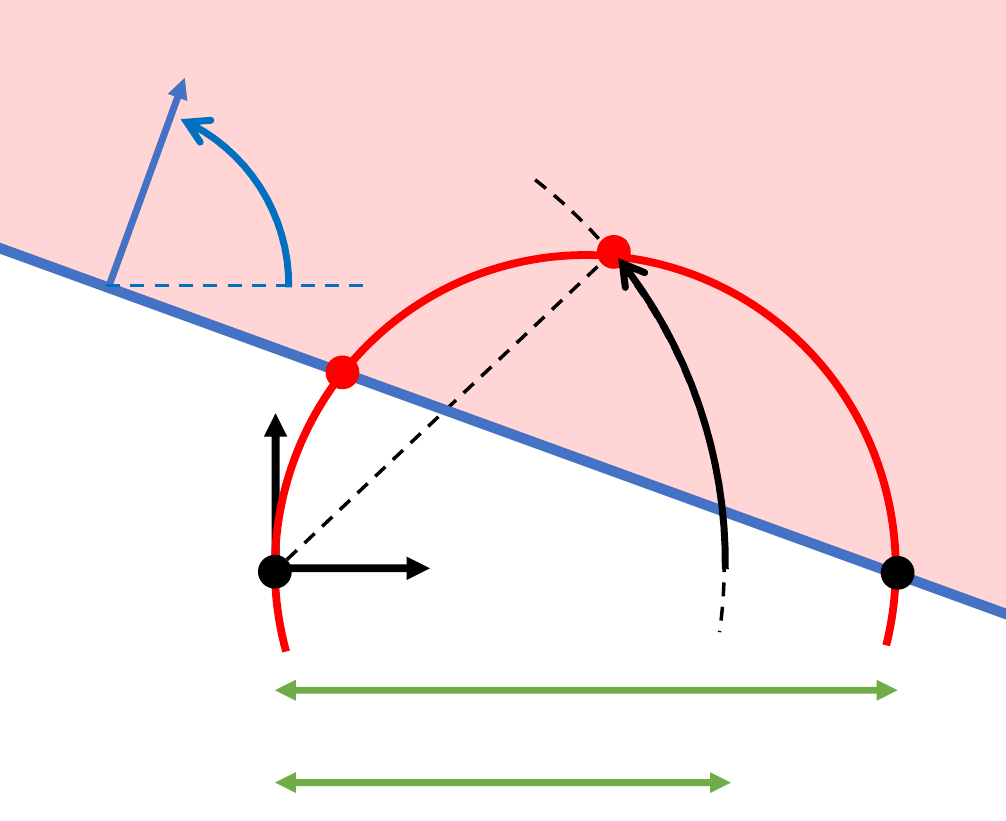}
\caption{The geometrical configuration of the problem in $\Plan$.}
\label{fig:config}
\end{figure}

\begin{prop}
Consider the mid-point $\c = (\x_o + \x_b)/2$.
The locus of the points $\z^\star(\theta)\in\Plan$ is the circle of center $\c$ and radius $d/2$.
Indeed, $\z^\star(0) = \x_b$ and $\z^\star(\pm\pi/2)=\x_o$.
\end{prop}
Little algebra shows that $\|\z^\star(\theta) - \c\| = d/2, \forall \theta\in[-\pi/2,\pi/2]$.
This circle is depicted in red in Fig.~\ref{fig:config}.


\begin{prop}
If $\z^\star(\theta)$ is adversarial, then so is $\z^\star(\phi)$ for $\phi\in[0,\theta]$. Conversely, if $\z^\star(\theta)$ is not adversarial, then so is $\z^\star(\phi)$ for $\phi\in[\theta,\mathsf{sign}(\theta).\pi/2]$. 
\end{prop}
This is due to the monotonicity of function $\alpha\to g_\alpha(\theta^\star(\alpha))$.

\begin{prop}
\label{prop:p3}
$\theta^\star = \psi$ is the angle yielding a maximum distortion decrease of $\alpha=1-\cos(\psi)$.
The point $\z^\star(\theta^\star)$ is indeed the projection of $\x_o$ on the boundary line $\dout\cap\Plan$:
$\z^\star(\theta^\star) = d\cos(\psi)\n + \x_o$. 
\end{prop}
This is shown by injecting~\eqref{eq:gAlpha} in~\eqref{eq:CondAdv}.

\subsection{Iterations over orthonormal directions}
This section assumes that the boundary $\dout$ is an affine hyperplane passing through $\x_{b,1}$ in $\real^\dim$,
with normal vector $\N$. We consider a random basis of $\text{span}(\x_{b,1} - \x_o)^\bot$ composed of $\dim-1$ vectors $\{\t_i\}_{i=1}^{\dim-1}$.
The normal vector is decomposed in spherical coordinates:
\begin{eqnarray}
\N &=& \sin(\psi_{D-1})\t_{\dim-1} + \cos(\psi_{\dim-1})\sin(\psi_{\dim-2})\t_{\dim-2}+\nonumber\\
&\ldots& + \cos(\psi_{\dim-1})\ldots\cos(\psi_2)\n_1,
\end{eqnarray}
where $\n_1 := \sin(\psi_1)\t_1 + \cos(\psi_1)\u_1$ is the $\ell_2$ normalized projection of $\N$ onto hyperplane $\Plan_1$ spanned by 
$\t_1$ and $\u_1 := (\x_{b,1} - \x_o)/d$. Note that $\N^\top\u_1 = \cos(\psi_{D-1})\ldots \cos(\psi_1)$.

We then find $\x_{b,2} := \z^\star(\theta^\star)\in\out\cap\Plan_1$ as given in Prop.~\ref{prop:p3} and define $\u_2 := (\x_{b,2} - \x_o) / d\cos(\psi_1)=\n_1$.
We solve the problem in $\Plan_2$ spanned by $\t_2$ and $\u_2$.
Note that $\N^\top\u_2 = \cos(\psi_{D-1})\ldots \cos(\psi_2) \geq \N^\top\u_1$.

\begin{prop}
Iterating this process converges to the adversarial point with minimal distortion.
\end{prop}
Iterations increase the scalar product between $\N$ and $(\x_{b,k}-\x_o) \propto \u_k $  given by:
\begin{equation}
\N^\top \u_k = \prod_{i=1}^{D-k} \cos(\psi_{D-i}).
\end{equation}
At the end, $\x_{b,D}\in\out$ and $\x_{b,D}-\x_o$ is colinear with $\N$, thus pointing to the projection of $\x_o$ to the hyperplane boundary.   

A clever strategy browses the directions according to the decreasing order of their angles $(|\psi_k|)_k$ (biggest distortion decreases first).
This is out of reach for the attacker oblivious to $\N$ and not willing to spend queries for its estimate.

\subsection{Convex boundary}
Our procedure can be seen as a coordinate descent on a random basis.
If the boundary $\dout$ is not a hyperplane but a smooth and convex surface, then cycling over the vectors $\{\t_i\}_{i=1}^{D-1}$ multiple times ensures convergence to a local minimum~\cite{Nesterov:2012aa}. On one hand, this reference shows that the rate of convergence of the random coordinate descent (on expectation) is essentially the same as the \emph{worst-case} rate of the standard gradient descent (when it is available). On the other hand, estimating the gradient in the black-box setting costs more queries than the coordinate descent of Sect.~\ref{ssec:BasicIdea}. These conflicting arguments deserve investigation.

\section{The \name\ attack}
\label{ssec:algo}
This section presents the attack based on the ideas explained in Sect.~\ref{sec:Approach}.
One iteration of \name is summarized in pseudo-code Alg.~\ref{alg:global_algo}.

\subsection{The algorithm}

\textbf{Initialisation}.
The algorithm needs an initial point $\x_{b,1}\in\dout$.
It first generates a point $\y_0\in\out$.
As done in~\cite{Rahmati:2020aa,Li:2020aa}, $\y_0$ is one image from the targeted class (targeted attack) or a noisy version of $\x_o$ (untargeted attack).
Defining $\y_\lambda = \lambda\x_o  + (1-\lambda)\y_0$, a binary search over $\lambda\in(0,1)$ results in $\x_{b,1}$ adversarial and close to the boundary.
  
 \textbf{New direction}.
At iteration $k$, the point $\x_{b,k}\in\out$ and close to the boundary defines $\u_k \propto \x_{b,k} - \x_o$, $\|\u_k\| = 1$.
Line~\ref{alg:draw} generates pseudo-randomly $\tt_k\sim\mathcal{T}$ (see Sect.~\ref{ssec:Distrib}).
A Gram-Schmidt procedure makes it orthogonal to $\u_k$ and to the $L$ (at most) last directions $\V_{k-1} := \{\t_j\}_{j = \max(k-L,1)}^{k-1}$, producing the new directions $\t_k$ in line~\ref{alg:GS}.
  
\textbf{Sign Search}. The algorithm considers points $\z(\alpha,\theta)$ as defined in~\eqref{eq:Z}
with $\u=\u_k$, $\t=\t_k$, $d_k:=\|\x_{b,k}-\x_o\|$, and the coupling $\cos(\theta) = 1-\alpha$.
The sign of $\theta$ depends on the sign of unknown $\psi$ (see Sect.~\ref{ssec:BasicIdea}).
Hence, we test some angles starting with the biggest amplitudes, alternating + and - sign, as stored in the vector
$\theta_{\max}.\taub$ with $\taub:=\left(1, -1, (T-1)/T, -(T-1)/T, \ldots,1/T, -1/T\right)$.

The search stops as soon as an adversarial image is found.
If this fails, line~\ref{alg:Decay} decreases $\theta_{\max}$, direction $\t$ is given up (line~\ref{alg:GiveUp}), and another direction is generated.

\textbf{Binary Search}. When the sign search finds an adversarial image at $\theta = \theta_{\max}.t/T$,
the binary search in line~\ref{alg:BS} refines the angle $\theta$ over the interval $\theta_{\max}/T.[t,t+\mathsf{sign}(t)]$ within $\ell$ steps.
The result is $\theta^\star$ s.t. $\z^\star(\theta^\star) \in\out$ is the new boundary point $\x_{b,k+1}$ provoking a distortion decrease $\alpha^\star = 1-\cos(\theta^\star)$.

\subsection{Distribution of the directions}
\label{ssec:Distrib}
The algorithm is a random process as it draws directions from distribution $\mathcal{T}$ according to Alg.~\ref{alg:dimension_redution_dct}.
This has two roles: dimension reduction and adaptivity to the content of $\x_o$.

Dimension reduction is implemented with the help of a reversible image transformation ($\DCT$ $8\times8$, or full frame $\DCT$ in Table~\ref{table:dct}).
Line~\ref{alg:select} selects a fraction $\rho$ of the transform coefficients, typically in the low frequency subband.
We draw $\rho\dim$ samples uniformly distributed over $\{-1,0,1\}$, the other transform coefficients being set to 0. The inverse transform yields the direction $\tt$ in the pixel domain.   

Adaptivity to the visual content makes the perturbation less perceptible thanks to the masking effect well know in image watermarking~\cite{Earl:2007aa}.
It shapes the adversarial perturbation like the visual content of $\x_o$. The following is a simple implementation of this principle: denote the $i$-th transform coefficient of image $\x_o$ by $X_{o,i}$. Line~\ref{alg:modul} modulates the amplitude of a random variables by  $A(|X_{o,i}|)$,
where $A:\real^+\to\real^+$ is a non decreasing function.
The goal is to shape the power distribution of the perturbation as the one of the original image.

\subsection{Interpolation}
\label{ssec:Interpolation}
\begin{figure}[tb]
\centering
\resizebox{\columnwidth}{!}{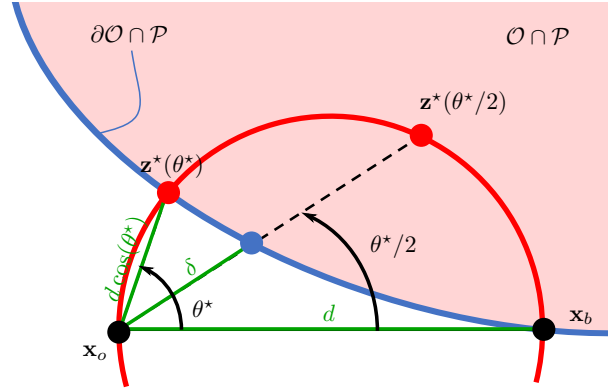}
\caption{Interpolation mechanism to refine the boundary point.}
\label{fig:Earl}
\end{figure}

Section~\ref{sec:Approach} motivated our design assuming the boundary is an hyperplane.
This extra interpolation is an \emph{option} of \name inspired by the watermarking attack~\cite{Earl:2007aa}, which tackles convex surfaces with small curvature as in Fig.~\ref{fig:Earl}.

A given iteration starts with $\x_{b,k}\in\dout$ at angle $\theta=0$ and distance $d$.
The binary search in line~\ref{alg:BS} gives the angle $\theta^\star$ of a boundary point at distance $d\cos(\theta^\star)$.
This option finds a third point on the boundary at angle $\theta^\star/2$ thanks to a binary search between $\x_o$ and $\z^\star(\theta^\star/2)$.
This point, depicted in blue in Fig.~\ref{fig:Earl}, is at distance $\delta \leq d\cos(\theta^\star/2)$.

Thanks to these three boundary points resp. at angle $0$, $\theta^\star/2$, and $\theta^\star$, we interpolate the mapping from angle to distance (of the surjection of $\z(\alpha,\theta)$ onto the boundary) by a second order polynomial and find its minimum at:
\begin{equation}
\hat{\theta} = \frac{\theta^\star}{4}\frac{4\delta - d(\cos(\theta^\star)+3)}{2\delta - d(\cos(\theta^\star)+1)}.
\end{equation}
This option concludes by a binary search finding the point on the boundary between $\x_o$ and $\z^\star(\hat{\theta})$.
The new point $\x_{b,k+1}$ is the closest point we found on the boundary.

\definecolor{myGreen}{RGB}{0,160,0}  
\def \da{\color{myGreen}d}
\def \db{\color{myGreen}\delta}
\def \dc{\color{myGreen}d\cos(\theta^\star)}

\def \proj{\mathsf{proj}}
\begin{algorithm}
\caption{One iteration of \name}
\label{alg:global_algo}
\begin{algorithmic}[1]
\Require {Original image $\x_o$, boundary point $\x_{b,k}\in\dout$, previous directions $\V_{k-1} := \{\t_j\}_{j = \max(k-L,1)}^{k-1}$}
\Ensure Output $\x_{b,k+1}\in\dout$, $\V_{k}$
        \State \underline{\textbf{New direction}}
        \State $\u_k = \eta(\x_{b,k} - \x_o)$ \Comment{$\eta(\x):=\x/\|\x\|$}
        \State $\tt_k\sim\mathcal{T}$  \Comment{Algorithm \ref{alg:dimension_redution_dct}}\label{alg:draw}
        \State $\t_k = \eta\left(\proj_{\mathsf{span}(\V_{k-1}\cup\u_k)^\bot}(\tt_k)\right)$ \Comment{Gram-Schmidt}\label{alg:GS}
        \State $\V_{k} = \V_{k-1}\cup\{\t_k\}$
        \State \underline{\textbf{Sign Search}}
        \State $j=1$, $\taub=\left(T, -T, (T-1), -(T-1), \ldots,1, -1\right)/T$
        \While{$\z^\star(\theta_{\max}.\tau_j)\notin\out \wedge j\leq 2T$}
        \State $j \leftarrow j+1$
        \EndWhile
        \If{$j < 2T$} \label{alg:Stop1}
        \State \underline{\textbf{Binary Search}}
        \State $\theta^\star = \mathsf{BS}(\theta_{\max}.\tau_j \,;\, \theta_{\max} (\tau_j+\mathsf{sign}(\tau_j)/T))$ \label{alg:BS}
        \State $\theta_{\max} \leftarrow \theta_{\max} / (1-\kappa)$
        \State Return $\x_{b,k+1} = \z^\star(\theta^\star)$
        \State or \underline{\textbf{Interpolation} Sect.~\ref{ssec:Interpolation}}
                \Else \Comment{Sign Search failed}
                \State $\theta_{\max} \leftarrow \theta_{\max}\times(1-\kappa)$ \Comment{Geometric decay}\label{alg:Decay}
        \State Go to line~\ref{alg:draw}  \Comment{Give up}\label{alg:GiveUp}
        \EndIf
        \end{algorithmic}
\end{algorithm}

\begin{algorithm}
\caption{Draw direction $\tt\sim\mathcal{T}$}
\label{alg:dimension_redution_dct}
\begin{algorithmic}[1]
\Require {Original image $\x_o$, frequency subband $\mathcal{F}$ s.t. $|\mathcal{F}| = \rho\dim$}, $A(\cdot)$ shaping function
\Ensure A random direction $\tt$ perceptually shaped as $\x_o$
    \State $\mathbf{X}_o = \DCT(\x_o)$
    \For{$j=1:n$}
    \If{$j\in\mathcal{F}$} \label{alg:select}
    \State $r\sim\mathcal{U}_{\{-1,0,1\}}$  \Comment $r\in\{-1,0,+1\}$
    \State $T_j = A(|X_{o,j}|)\times r$ \label{alg:modul}
    \Else
    \State $T_j = 0$
    \EndIf
    \EndFor
    \State Return $\tt = \eta\left(\DCT^{-1}(\mathbf{T})\right)$ \Comment{$\eta(\x):=\x/\|\x\|$}
\end{algorithmic}
\end{algorithm}


\section{Experimental Work}
\label{sec:Experimental}

We first specify the experimental setup and the
parameters of our approach.  We then perform an ablation study on
\name (subsection \ref{ss:ab}), for it allows to precise gains
on the two considered metrics. Subsection~\ref{ss:perf}
performs a head-to-head comparison of all the competing approaches.

\subsection{Datasets and Experimental Setup}

\paragraph{Datasets}

For MNIST, we use a pre-trained CNN
network that is composed of 2 convolutional layers and
2 fully connected Layers.
Its accuracy is 99.14\%.
A subset of 100 \emph{correctly} classified images have been randomly
chosen to perform the ablation study.
Our attack generates directions on the pixel domain without any dimension reduction.

The ImageNet dataset is tackled by a pre-trained ResNet18,
made available for the PyTorch environment \cite{NEURIPS2019_9015}.
Its top-$1$ accuracy is $0.3024$.
We randomly selected 350 \emph{correctly} classified images from the ILSVRC2012's validation set with size $\dim = 3 \times 224 \times 224$.

\paragraph{Setup and Code}
We now detail the specific parameters of \name, for both MNIST and
ImageNet.  We set empirically the following values in Alg.~\ref{alg:global_algo}: $T=3$, $L=100$, $\theta_{\max} = 30$,
$\kappa=0.02$, at most $\ell = 10$ steps for the binary search (with an early stop if the range is lower than $1\%$ of $d$).
We develop \name on top of the FoolBox library.

\paragraph{Evaluation Metrics}
\def \q{\mathbf{q}}
The two core evaluation metrics are the
amount of queries, and the resulting
distortion on the attacked image.
The distortion is measured with the $\ell_2$ norm over the space $[0,1]^\dim$ (with $D$ the number of pixels times the number of colour channel).
For a given $\x_o$, it is the smallest distortion obtained over the sequence of queries $(\q_j)_{j=1}^k$ that happen to be adversarial:
\begin{equation}
d(k,\x_o) := \min_{1\leq j\leq k\,: \dec(\q_j)\neq\dec(\x_o)}\|\q_j - \x_o\|_2
\end{equation}
The mean over $N$ original images gives a characteristic of the attack efficiency
revealing its capacity to find an adversary close to the original image and especially its speed.
\begin{equation}
\label{eq:MeanDisto}
d(k):= \frac{1}{N}\sum_{i=1}^N d(k,\x_{o,i})
\end{equation}
We define the success rate as the probability of getting a distortion lower than a target $d_t$ within a query budget $K$:
\begin{equation}
\label{eq:SuccessRate}
S(d_t,K) := \frac{|\{i: d(K,\x_{o,i})\leq d_t\}|}{N}
\end{equation}

\subsection{Ablation Studies}
\label{ss:ab}
\def \ss {\texttt{SignSearch}}

\paragraph{Impact of the components - MNIST}
This first ablation evaluates how the hyperplane hypothesis \cite{Fawzi_2018_CVPR} meets a practical experimentation,
and how the interpolation mechanism of Sect.~\ref{ssec:Interpolation} is able to compensate this hypothesis.
To this end, four variants of our attack are tested in Fig.~\ref{fig:abl_stud_algo} and~\ref{fig:abl_stud_variance}:
\ss\ stops at line~\ref{alg:Stop1} of Alg.~\ref{alg:global_algo} whereas \name is the regular attack, `+Interpolation' uses the option~\ref{ssec:Interpolation}.

Our attack is highly random due to the generation of directions.
This may yield unstable results with adversarial images of scattered distortion.
Fig.~\ref{fig:abl_stud_algo} shows the distortion decrease averaged over $100$ images and Fig.~\ref{fig:abl_stud_variance}
the standard deviation for one image attacked 20 times.

This outlines the trade-off between the complexity of one iteration in terms of query number and the gain in the distortion decrease.
The Interpolation option may yield substantial decrease depending on the direction. This explains its large standard deviation.
Yet, its costs (2 more binary searches) slows down the speed.
\ss\ is less costly and offers competitive distortions only at the beginning. 
\name strikes the right trade-off both in term of averaged distortion and standard deviation.
Compared to \ss, it always exhausts the explored direction giving the best gain under the hyperplane boundary assumption.   
The first important insight is that this hypothesis seems to be good enough to ensure a rapid decay.

The ablation study also tested different values for some parameters of \name.
The value of $\kappa$ has no significant impact provided that $\kappa>0$.
Parameter $T$ doesn't benefit from higher value because of the finer seach in line~\ref{alg:BS}.


\def \ww{0.85\columnwidth}
\begin{figure}[t]
\centering
\resizebox{\ww}{!}{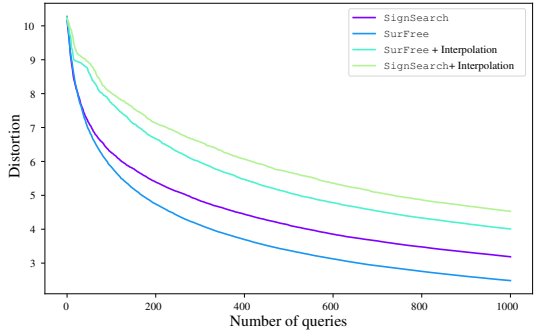}
\caption{Ablation study on \name. Mean distortion $d(k)$~\eqref{eq:MeanDisto} vs. number $k$ of queries on MNIST.}
\label{fig:abl_stud_algo}
\end{figure}



\begin{figure}[t]
\centering
\resizebox{\ww}{!}{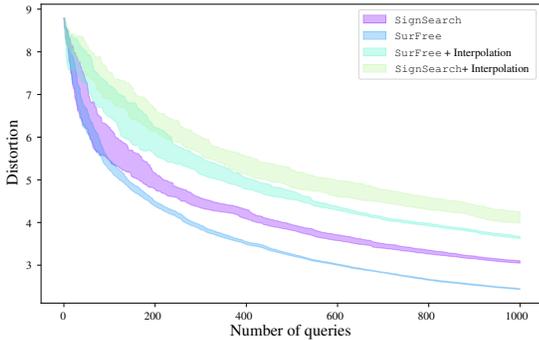}
\caption{Ablation study on \name. The deviation of the distortion over 20 runs of \name on one MNIST image.}
\label{fig:abl_stud_variance}
\end{figure}

\paragraph{Impact of the direction generation domain - ImageNet}
The literature reports that black-box attacks have difficulty in handling large images like ImageNet.
Attack become slow because the space is too large to be explored efficiently.
All competing attacks resort to a dimension reduction, typically by leveraging a
full $\DCT$ transform~\cite{Li:2020aa,Rahmati:2020aa}.
Yet, dimension reduction lowers the degrees of freedom for the attacker: the closest adversarial as defined in~\eqref{eq:OptDef} has a bigger distortion under this constraint. 
The distortion supposedly converge faster but to a bigger limit.
 
\name is no exception.
Table~\ref{table:dct} shows that the distortion in the full pixel domain is bigger within the first thousand queries.
For the same query budget, constraining the perturbation to lie in a smaller low-frequency subspace defined with the full $\DCT$ as in~\cite{Li:2020aa,Rahmati:2020aa}
is beneficial. Yet, this holds to some extent: distortion reported for a reduction of $\rho = 25\%$ are always larger than those for $50\%$. 

We now question the type of $\DCT$ transform.
Indeed, while the $\DCT$ full frame is widely acclaimed, we prefer the block-based $\DCT$ as used in JPEG.
It gives a better space-frequency localization trade-off.
Table~\ref{table:dct} shows that it does change the distortions a lot.
The last rows of Table~\ref{table:dct} focus on the adaptivity to the visual content of the original image (see Sect.~\ref{ssec:Distrib}).
Amplitude function $A(x) = x$ concentrates the perturbation power too much on some high amplitude coefficients when the original image has sharp edges.
$\tanh(x)$ is a good compromise between the constant and the identity functions.
It offers early distortion drop and reaches similar levels than $A(x) = cst$ in the long run. 


Our design is driven by the small query budget requirement so we choose $\tanh$ shaping function
applied on $\rho=50\%$ of the $\DCT_{8\times8}$ coefficients.

\begin{table}[t]
    \centering
\resizebox{\linewidth}{!}{%
\setlength\tabcolsep{2pt}
    \begin{tabular}{|c|c|c||c|c|}
    \hline
        Space & Shaping $A(x)$ & Dim. Reduc. $\rho$ & $K=100$  & $K=1000$ \\
        \hline\hline
            Pixel & \_ & 100\% & 27.23 & 17.20 \\
        \hline
        \hline
            $\DCT_{full}$ & $cst$ & 50\% & 26.50 &  15.35 \\
            $\DCT_{full}$ & $cst$ & 25\% & 32.08 &  21.23 \\
        \hline
        \hline
        	    $\DCT_{8\times8}$ & $cst$ & 50\% & 19.49 & 10.69 \\
	    $\DCT_{8\times8}$ & $cst$ & 25\% & 18.26  & \textbf{9.93} \\
        \hline
            $\DCT_{8\times8}$ & $x$ & 50\% & 20.11 &  11.96 \\
            $\DCT_{8\times8}$ & $x$ & 25\% & 20.29 &  12.22 \\
        \hline
	    $\DCT_{8\times8}$ & $\tanh(x)$ & 50\% & \textbf{17.38} & 10.22 \\
            $\DCT_{8\times8}$ & $\tanh(x)$ & 25\% & 18.20 & 10.61 \\
            \hline
    \end{tabular}
}
\caption{Mean distortion $d(K)$ when random directions are generated with different subspaces and shaping (ImageNet).}
\label{table:dct}
\end{table}


\subsection{Benchmarking}
\label{ss:perf}

We compare to recent algorithms considered as state-of-the-art decision-based black-box attacks: \hsja, \geoda and \qeba.
These 3 algorithms leverage gradient surrogates.
The benchmark does not include older attacks like \texttt{OPT}~\cite{Cheng:2019aa} and \texttt{BA}~\cite{Brendel:2018aa} because they have proven less efficient than the three above-mentioned references. 

\begin{figure}[tb]
\centering
\resizebox{\ww}{!}{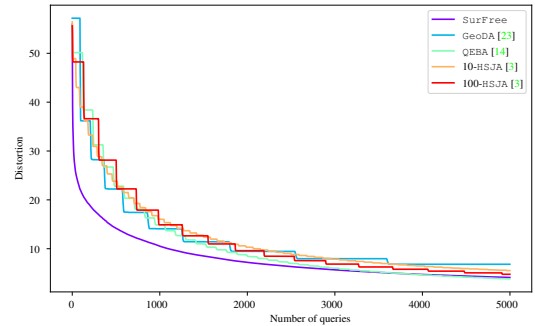}
\caption{Benchmark on ImageNet. The amount of queries $k$ ($x$-axis) w.r.t. mean distortion $d(k)$ ($y$-axis).}
\label{fig:Benchmark}
\end{figure}

We use the authors code for these algorithms: \hsja is integrated in
the FoolBox library \cite{rauber2017foolbox, rauber2017foolboxnative}.
For \geoda and \qeba,
we pull implementations from their respective GitHub
repositories\footnote{QEBA:
  https://github.com/AI-secure/QEBA}\footnote{GeoDA:
  https://github.com/thisisalirah/GeoDA} with default parameters.
For \geoda, the number of queries devoted to the gradient estimates follow a geometric progression of common ratio $\lambda^{-2/3}$ with $\lambda=0.6$, and the dimension reduction focuses on 5,625 coefficients of the full $\DCT$ transform.
Concerning \qeba, $\rho=25\%$ dimension reduction on low frequency full $\DCT$ coefficients.
\hsja works on the pixel domain, the number of queries devoted to gradient estimates scales as $N_0\sqrt{j}$ with $j$ the iteration number.
We tested two versions with $N_0\in\{10,100\}$, which is directly observable with the larger plateaus on Fig.~\ref{fig:Benchmark}.

A very important point is that all attacks are initialized with 
the same first adversarial example in order to avoid favoring a competitor by giving it an easier initialization.


\begin{table*}[t!]
	\centering
	\resizebox{\linewidth}{!}{%
	\setlength\tabcolsep{2pt}
		\begin{tabular}{|c||cccc||cccc||cccc|}
		\hline
			{target} & \multicolumn{4}{c||}{$K=500$ queries} & \multicolumn{4}{c||}{$K=1,000$ queries}  & \multicolumn{4}{c|}{$K=2,000$ queries} \\
			$d_t$ & \hsja & \geoda & \qeba & \name & \hsja & \geoda & \qeba & \name & \hsja & \geoda & \qeba &
			 \name  \\
			 \hline
			 \hline
			 30 & 0.56 & 0.79 & 0.71 & \textbf{0.90} & 0.88 & 0.93 & 0.88 & \textbf{0.96} & 0.98 & 0.96 & 0.97 & \textbf{0.99} \\
			 10 & 0.13 & 0.25 & 0.32 & \textbf{0.44} & 0.23 & 0.52 & 0.46 & \textbf{0.57} & 0.40 & 0.70 & 0.69 & \textbf{0.73} \\
			 5 & 0.07 & 0.14 & 0.17 & \textbf{0.23} & 0.09 & 0.21 & 0.30 & \textbf{0.31} & 0.13 & 0.39 & 0.47 & \textbf{0.50} \\
			 \hline
		\end{tabular}
	}
	\caption{Success rate $S(d_t,K)$ for achieving a targeted distortion $d_t$ under a limited query budget $K$ (ImageNet).}
	\label{table:accuracy_distortion}
\end{table*}

\def \wa{0.1\linewidth}
\def \ad{\tiny{amer. dipper}}
\def \si{\tiny{siamang}}
\def \sw{\tiny{stone wall}}
\def \bt{\tiny{terrier}}
\def \st{\tiny{stingray}}
\def \me{\tiny{megalith}}
\def \gu{\tiny{guenon}}
\def \ti{\tiny{titi monkey}}
\def \pt{\tiny{ptarmigan}}
\def \hu{\tiny{hummingbird}}
\def \cd{\tiny{cliff dwelling}}
\def \tr{\tiny{triceratops}}
\def \wo{\tiny{wombat}}
\def \ar{\tiny{armadillo}}
\def \tu{\tiny{tusker}}
\def \br{\tiny{brambling}}

\begin{table*}[t]
\centering
\resizebox{\linewidth}{!}{%
{\def\arraystretch{0.85}
\setlength\tabcolsep{1pt}
\begin{tabular}{|c|ccc|ccc|ccc|}
\hline
attack & $K = 100$ & $K = 500$ & $K= 1000$ & $K = 100$ & $K = 500$ & $K= 1000$ & $K = 100$ & $K = 500$ & $K= 1000$ \\
\hline\hline
\name &
\includegraphics[trim = 113  36 101 35,clip,width=\wa]{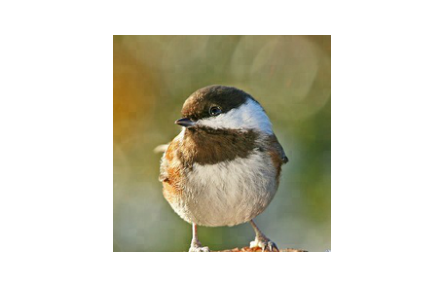} &
\includegraphics[trim = 113  36 101 35,clip,width=\wa]{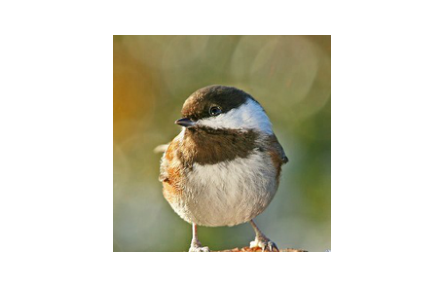} &
\includegraphics[trim = 113  36 101 35,clip,width=\wa]{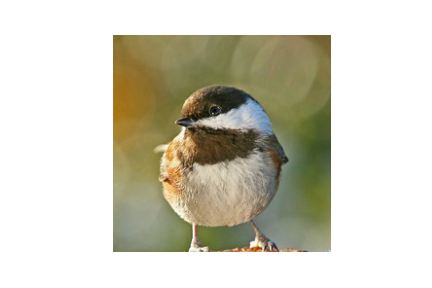} &
\includegraphics[trim = 113  36 101 35,clip,width=\wa]{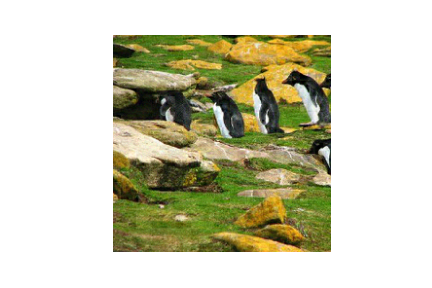}&
\includegraphics[trim = 113  36 101 35,clip,width=\wa]{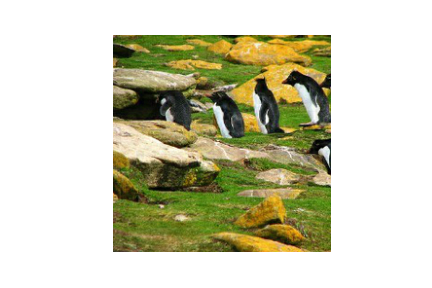}&
\includegraphics[trim = 113  36 101 35,clip,width=\wa]{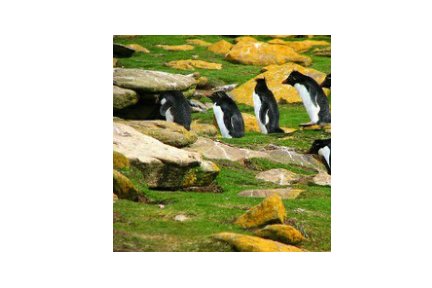}&
\includegraphics[trim = 113  36 101 35,clip,width=\wa]{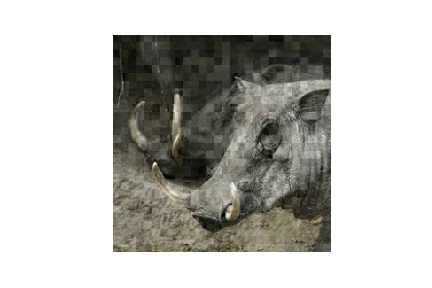} &
\includegraphics[trim = 113  36 101 35,clip,width=\wa]{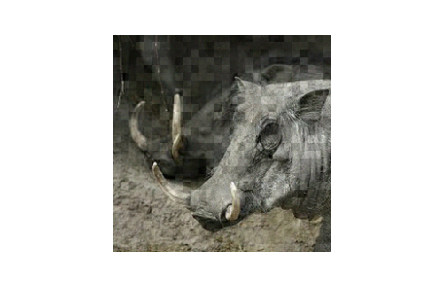} &
\includegraphics[trim = 113  36 101 35,clip,width=\wa]{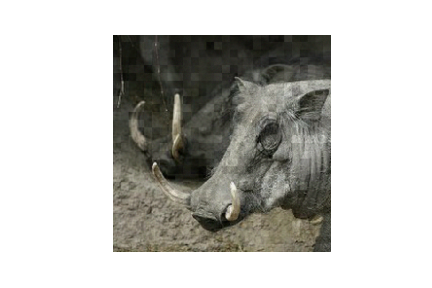}\\
 & \ad - 2.6 & \ad - 1.3 & \ad - 0.9 & \sw - 14.9 & \sw - 8.7 & \sw - 5.4 & \cd- 21.9 & \cd - 18.4 & \tr - 13.5 \\
 \hline 
\qeba &
\includegraphics[trim = 113  36 101 35,clip,width=\wa]{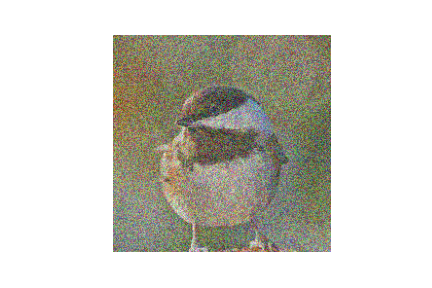} &
\includegraphics[trim = 113  36 101 35,clip,width=\wa]{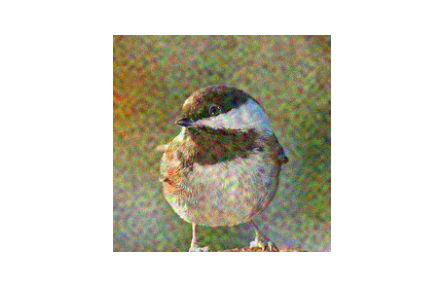} &
\includegraphics[trim = 113  36 101 35,clip,width=\wa]{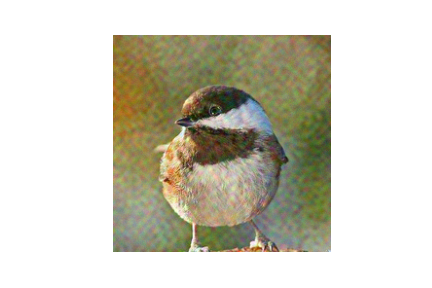}&
\includegraphics[trim = 113  36 101 35,clip,width=\wa]{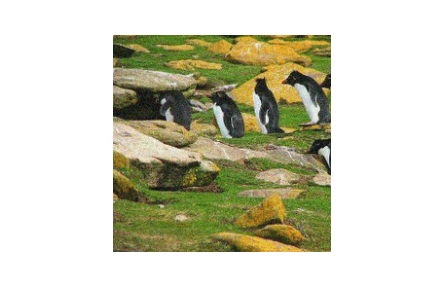} &
\includegraphics[trim = 113  36 101 35,clip,width=\wa]{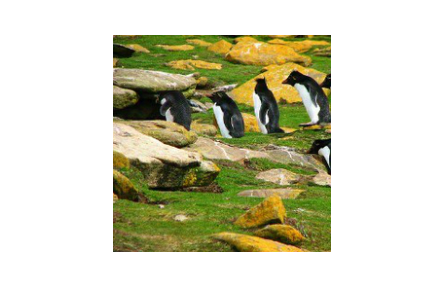} &
\includegraphics[trim = 113  36 101 35,clip,width=\wa]{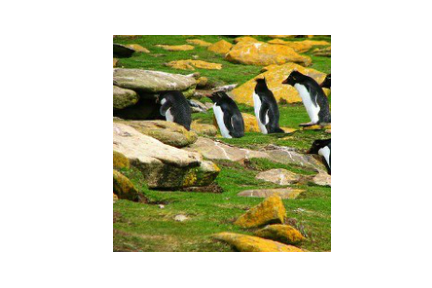} &
\includegraphics[trim = 113  36 101 35,clip,width=\wa]{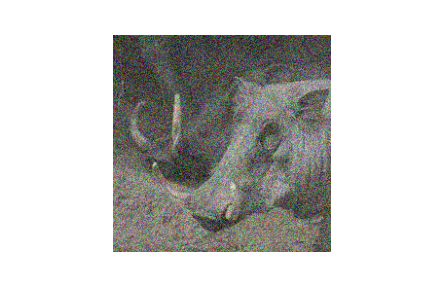} &
\includegraphics[trim = 113  36 101 35,clip,width=\wa]{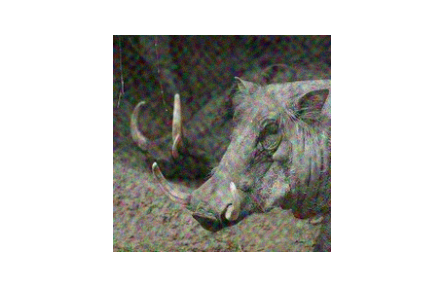} &
\includegraphics[trim = 113  36 101 35,clip,width=\wa]{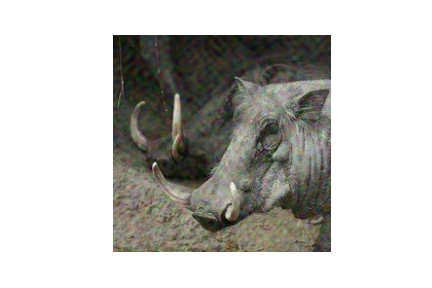}\\
& \st - 60.6 & \st - 33.7 & \st - 20.8 & \sw - 25.2& \sw - 4.8 & \sw - 2.6 & \wo - 58.3 & \wo - 24.3 & \wo - 13.6 \\
\hline
\geoda &
\includegraphics[trim = 113  36 101 35,clip,width=\wa]{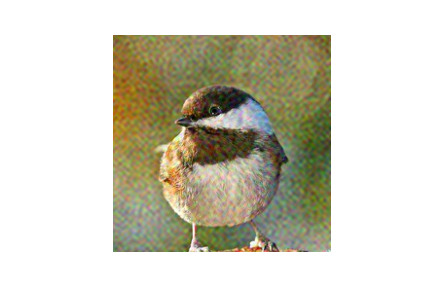}&
\includegraphics[trim = 113  36 101 35,clip,width=\wa]{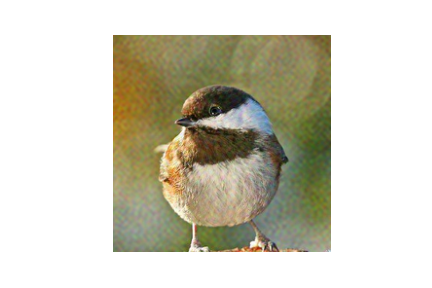}&
\includegraphics[trim = 113  36 101 35,clip,width=\wa]{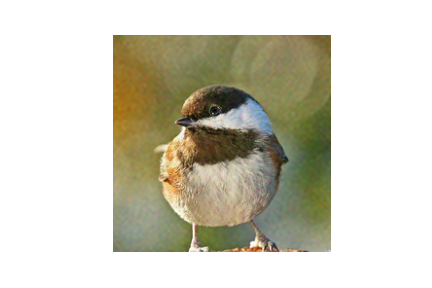}&
\includegraphics[trim = 113  36 101 35,clip,width=\wa]{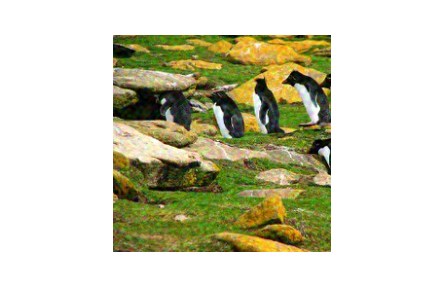}&
\includegraphics[trim = 113  36 101 35,clip,width=\wa]{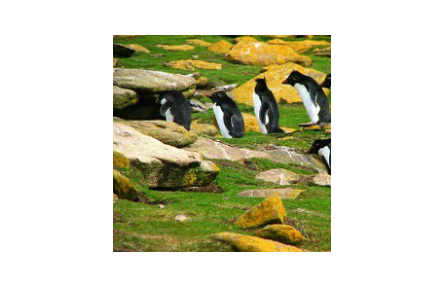}&
\includegraphics[trim = 113  36 101 35,clip,width=\wa]{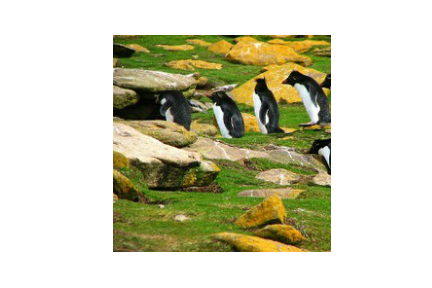}&
\includegraphics[trim = 113  36 101 35,clip,width=\wa]{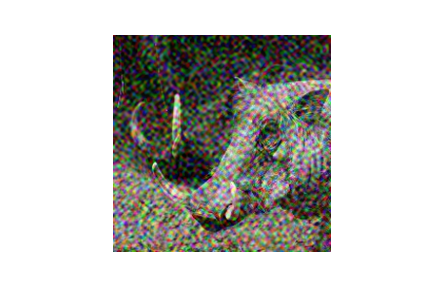}&
\includegraphics[trim = 113  36 101 35,clip,width=\wa]{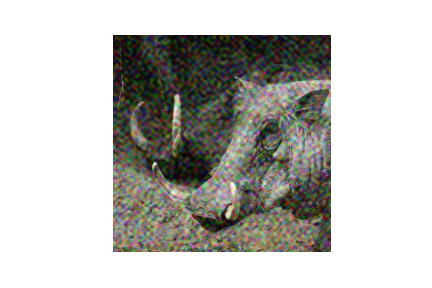}&
\includegraphics[trim = 113  36 101 35,clip,width=\wa]{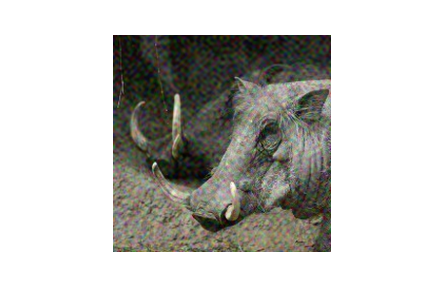}\\
& \br - 18.9 & \br - 9.7 & \br - 5.8 & \sw - 15.8 & \me - 4.5 & \me - 2.6 & \ar - 49.4 & \tu - 31.3 & \tu - 18.9  \\
\hline
\end{tabular}
}
}
\caption{Visual trajectories for an easy (chickadee), a medium (king penguin), and a difficult image (warthog) - predicted label and distortion}
\label{tab:visual}
\end{table*}

\begin{figure}[b]
\centering
\resizebox{\ww}{!}{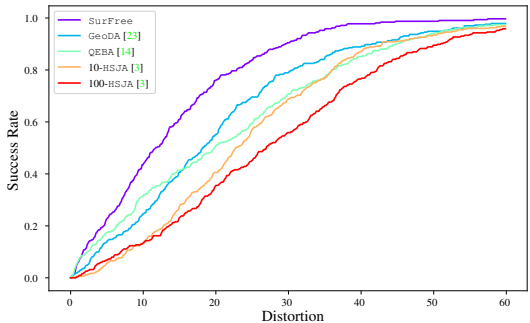}
\caption{Success rate $S(d_t,K)$~\eqref{eq:SuccessRate}
 vs. target distortion $d_t$ with $K = 500$ queries over ImageNet.}
\label{fig:Benchmark_distortion}
\end{figure}

\paragraph{Performance evaluation: distortion vs. queries}
Figure~\ref{fig:Benchmark} displays the distortion of the perturbation
($\ell_2$ norm) versus the amount of queries.  \name presents a smooth
curve, resulting from the averaging over 350 images.  The other
attacks still show large plateaus (as highlighted in
Fig.~\ref{fig:budget} for one image) because gradient estimates are
scheduled at the same instants for any image.  Note that these
plateaus are not shown in the papers because the distortion is seen as
a function of the iteration number, not the query number.  The two
most recent attacks, \qeba and \geoda indeed beat \hsja as reported in
the corresponding papers.  \name dives significantly faster than
all attacks to lower distortions (most notably from 1 to 750
queries), while \qeba prevails at around $3,750$ queries.
Note that \name is also first with $\DCT$ full but for a shorter period ($\approx 800$ queries).
For completeness, here are the scores at 10,000 queries:
2.09 (\qeba) $<$ 2.72 (\name) $<$ 3.48 (HSJA\_10) $<$ 4.63
(\geoda). Although a small query budget drives its design,
\name is not off in the long run.
Similar results are observed for MNIST (in the pixel domain, without dimension reduction) where \name is ahead up to $\approx 5,000$ queries.

\paragraph{Performance evaluation: Success rate}
We now consider three query budgets, $K\in\{500,1,000, 2,000\}$, which are rather low with regards to the state-of-the-art (see Sect.~\ref{ssec:BB}).

Table~\ref{table:accuracy_distortion} details how the success rate $S(d_t,K)$ varies for some setup $(d_t,K)$~\eqref{eq:SuccessRate}.
Fig.~\ref{fig:Benchmark_distortion} shows the success rate $S(d_t,500)$ increase with $d_t$.
\geoda is superior to \qeba for large target distortions only.
Both schemes outperform \hsja.
\name remains the best attack for any target distortion up to this $2,000$ query budget. 

Finally, Table~\ref{tab:visual} displays the visual trajectories of
three attacked images witnessed as easy, medium, and
difficult to attack for \name. While all three attacks affect differently the images, \name gives relatively less annoying artefacts.
We also note a drawback of \qeba: the adversarials often keep the label of the random starting point (\eg stingray), hence sometimes converging to a local minimum which is far from the optimal solution~\eqref{eq:OptDef}.
%


\section{Conclusion}
\label{sec:Conclusion}

The performance of black box decision-based attacks reveals important
gaps when it comes to the required amount of queries.
Core to the three state-of-the-art approaches this papers considers is the
estimation of gradients. This step is particularly costly, with
regards to our novel geometrical attack \name.
The trial of multiple directions together with a simple mechanism getting the best distortion decrease along a given direction 
allow a fast convergence to qualitative adversarials, within an order of
hundreds of queries solely. This sets a new stage for future works.


\newpage
{\small
\bibliographystyle{ieee_fullname}
\bibliography{biblio}
}

\end{document}